\documentclass[final]{siamltex}

\usepackage{citesort}
\usepackage[dvips]{graphicx}

\newcommand{\Figref}[1]{Fig.~\ref{#1}}
\newcommand{\Eqref}[1]{Eq.~(\ref{#1})}

\title{
Fractal Basins of Attraction Associated with a
Damped Newton's Method
\thanks{This research was partially supported by grants
from NSF NSF-DMS-93-07893 and DOE-DE-FG05-94ER25214}
}

\author{Bogdan I. Epureanu \thanks{Department of
Mechanical Engineering, Duke University, Box 90302,
Durham, NC, 27708-0302 { \tt bie@me1.egr.duke.edu} }
\and Henry S. Greenside \thanks{Department of Computer
Science, Duke University, Box 90129, Durham, NC,
27708-0129 { \tt hsg@cs.duke.edu} } }

\begin{document}

\maketitle

\begin{abstract}
  An intriguing and unexpected result for students
  learning numerical analysis is that Newton's method,
  applied to the simple polynomial~$z^3 - 1 = 0$ in the
  complex plane, leads to intricately interwoven basins
  of attraction of the roots. As an example of an
  interesting open question that may help to stimulate
  student interest in numerical analysis, we
  investigate the question of whether a damping method,
  which is designed to increase the likelihood of
  convergence for Newton's method, modifies the fractal
  structure of the basin boundaries. The overlap of the
  frontiers of numerical analysis and nonlinear
  dynamics provides many other problems that can help
  to make numerical analysis courses interesting.
\end{abstract}



\pagestyle{myheadings}

\thispagestyle{plain}

\markboth{B. Epureanu and H. Greenside}{Fractal Basins of Attraction}

\section{Introduction}

Numerical analysis is an immensely exciting area of
research with many intellectual and practical benefits
and yet this excitement sometimes doesn't come across
in courses and textbooks.  One reason is that courses
and texts sometimes do not indicate some of the
frontiers of the field and why these frontiers are
interesting. Perhaps more importantly, few introductory
courses and texts make students aware that they are
actually close to, and can contribute to, research
frontiers even while participating in an introductory
course. The ease of finding and exploring new problems
in numerical analysis is one of its particular
attractions.

\begin{figure}[ht]  
  \centering
  \includegraphics[width=3.5in]{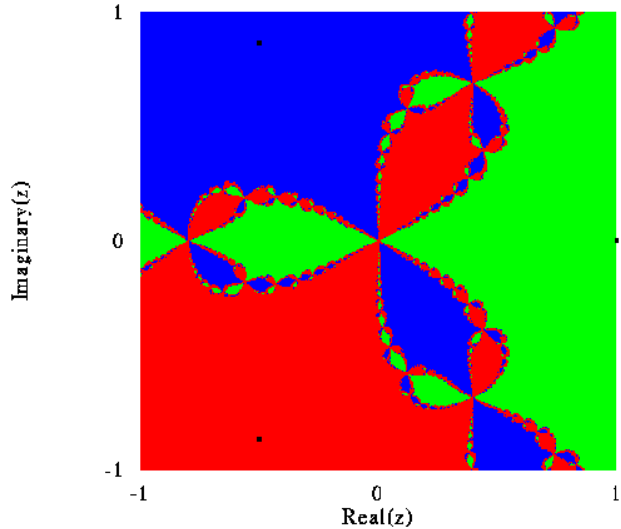}
\caption{The regions of three different colors indicate
  the approximate basins of attraction for the three
  roots of the complex polynomial~$z^3 - 1 = 0$ in the
  square region defined by $-1 \le Real(z) \le 1$ and
  $-1 \le Imag(z) \le 1$. The positions of the roots of
  unity are indicated by the small black squares. This
  figure was computed numerically by using a mesh of
  $1000 \times 1000$ pixels inside this square. The
  center point of each pixel was used to start a Newton
  iteration under \Eqref{eq:Newton-mapping}. The pixel
  was assigned a particular basin color if convergence
  to a root was identified within 1000~iterations. The
  criterion for convergence was that both the
  magnitudes of the correction~$z_{i+1} - z_i$ and of
  the residual $z_i^3 - 1$ were smaller than the
  value~$10^{-12}$.  The boundaries separating the
  basins are intricately braided, with a fractal
  dimension of~$D \approx 1.42$.  }
\label{fig:Newton} 
\end{figure}

As an illustration of this point that we hope will be
of pedagogical interest to those teaching and thinking
about numerical analysis, we pose and partially solve a
question that was raised by the attractive cover
picture of the book by Kincaid and Cheney
\cite{KincaidCheney96}. (This book has been used at
Duke University for the last few years in its graduate
introductory course on numerical analysis; the first
author was a recent student in this course when it was
taught by the second author.) This cover picture is
reproduced in \Figref{fig:Newton}
and reveals a remarkable result that is now nearly
120~years old \cite{Cayley1879,Nauenberg89,Crownover95}: 
for the simple polynomial equation~$z^3 - 1 = 0$ in the 
complex plane, the
sets of points~$z_0$ that converge to a root under the
map given by Newton's method,
\begin{equation}
  \label{eq:Newton-mapping}
  z_{i+1} = z_i  -
    \left({  z_i^3 - 1 }  \over  3 z_i^2 \right) .
\end{equation}
have an unexpected intricate geometric complexity. In
the language of nonlinear dynamics \cite{Strogatz94},
the set of points that converge towards a fixed point
under successive iterations of some mapping is called
the {\em basin of attraction} of that fixed
point. \Figref{fig:Newton} then says graphically that
the basins of attraction for Newton's method are
bounded by intricately braided objects. Naively, one
would have expected the basins of attraction to be
given by the sets of points closest to a given root and
so bounded by straight lines.

This braided structure is typical of geometric objects
known as fractals \cite{Strogatz94,Crownover95}. The
concept of a fractal is hard to define rigorously
\cite{Mandelbrot83} but can be casually defined as a
scale-invariant set that has a non-integer
dimension~$D$ known as its fractal dimension. For
\Figref{fig:Newton}, scale invariance corresponds to
the fact that if any small region of the braided
structure is greatly magnified, further braided
structure is found of a similar complexity and there is
no end to this intricate detail upon further and
further magnification. The fractal dimension is perhaps
best understood as a scaling exponent that generalizes
the usual concept of dimension to non-integer values
\cite{Strogatz94}. E.g., one kind of fractal dimension
known as the capacity~$C$ indicates the
scaling~$N_\varepsilon \propto \varepsilon^{-C}$
as~$\varepsilon \to 0$ of the minimum number of
balls~$N_\varepsilon$ of radius~$\varepsilon$ needed to
cover the set as the ball radius~$\varepsilon$ tends to
zero\footnote{More precisely, we would define the
capacity as the limit~$C =\lim_{\varepsilon \to 0} \log
N(\varepsilon) / \log(1/\varepsilon)$ if this limit
exists. In practice, one computes numerically a
sequence of values~$N_{\varepsilon_i}$ for different
radii~$\varepsilon_i$ and then estimates the dimension
from the slope of a least-squares linear fit to a plot
of~$\log(N_\varepsilon)$
versus~$\log(1/\varepsilon)$.}.  Numerical calculations
of the capacity of a set are often slowly convergent
when the points of the set are visited with non-uniform
probability (measure) by the dynamics
\cite{Greenside82pra}. For this reason, in the
following we have used the average pointwise
dimension~$D$ \cite{Moon92} to quantify the structure
of the fractal basin boundaries. This dimension takes
into account the probability of visiting different
parts of a set and so is more rapidly convergent in
practice.\footnote{The point-wise dimension~$D$ is
defined by the limit ~$D =\lim_{\varepsilon \to 0} \log
P(\varepsilon) / \log(1/\varepsilon)$ if this limit
exists, where the quantity~$P(\varepsilon)$ is the
density or probability of points in a ball of
radius~$\varepsilon$, i.e., the ratio of the number of
points in the ball to the total number of points in the
set.}.

Another interesting aspect of the fractal structure in
\Figref{fig:Newton} is that the boundaries satisfy the
counterintuitive Wada property \cite{Nusse96}: every
point on the boundary of one basin of attraction is
also on the boundary of the other basins of attraction.

The Kincaid and Cheney book touches only briefly on
these remarkable facts and does not mention that
fractal basins are in fact typical for Newton methods,
rather than rare or pathological. Some of the important
(and largely numerical) discoveries in the field of
nonlinear dynamics in the last twenty years is that
fractal structures pop up over and over again for
nonlinear equations and for typical choices of equation
parameters and of initial conditions: basin boundaries
can be fractal, the attracting limit set for maps,
flows, and partial differential equations can be
fractal, and the set of parameter values that lead to a
particular class of solutions (e.g., fixed points) can
be fractal \cite{Strogatz94}. This fractal structure
has important practical implications for many numerical
problems and algorithms and so numerical analysis
students should be aware of this. As one example,
fractal basins of attraction for Newton's method imply
that it will generally be difficult or impractical to
characterize the ever-so-important initial guesses that
will lead to convergence. In some extreme cases, it may
be impossible to characterize the dependence of the
algorithm or even of a laboratory experiment on
parameters \cite{Ott94}.

\Figref{fig:Newton} is a good example of a result that
is easily computed by students in an introductory
numerical analysis course and that leads to
interesting, and largely unexplored, questions that
might entice a student to get interested in numerical
analysis. Thinking about this figure carefully raises
the following kinds of questions: Are fractal
boundaries associated with Newton's method rare or
common? What determines the braided geometric structure
(e.g., its fractal dimension~$D$)?  Do other
root-finding algorithms lead to such fractal structure
and, if so, can one relate the details of the algorithm
to the fractal geometry? How do floating point errors
affect the fractal structure? What are the implications
of these fractal basin boundaries for root-finding
software?

In the following, we address briefly one of the above
questions, namely do different root-finding
algorithms---particularly Newton's method with
damping---also lead to fractal basins of attraction? A
well known weakness of Newton's method is that it is
locally, but rarely globally, convergent
\cite{KincaidCheney96}. To improve this situation,
researchers have invented damping methods
\cite{Hager88,Dennis83} that try to increase the
likelihood that an initial condition will lead to
convergence. The basic strategy of most damping methods
is to add only a fraction of a Newton step when
correcting the current estimate of a root. During
successive iterations, the damping algorithm usually
increases the fraction of the Newton step towards the
value of one until full Newton steps are taken and
quadratic convergence is achieved.

A question which, to our knowledge, seems not to have
been discussed previously is then this: for fixed
points with fractal basins of attraction, do damping
methods change the fractal structure?  If so, do they
diminish or increase the fractal dimension~$D$ of the
boundaries? An interesting related question is whether
damping works by increasing the basin of attraction, in
that the basin with damping strictly contains the basin
without damping. An increase in the fractal
dimension~$D$ would possibly be an undesirable effect
since more initial conditions would lie near the
boundaries, making it harder to find criteria for
identifying good initial conditions, which is often the
most challenging part of a Newton method for many
scientific problems. The question of how damping may
modify the fractal basins of attraction is a natural
one to ask given material normally covered in an
introductory numerical analysis course and is also
within the ability of students to investigate
themselves.

\section{The Fractal Basins Associated with Armijo's Rule}

To study numerically and graphically how the basins of
attraction may be modified by damping, we consider one
of the simplest damping methods, the Armijo rule
\cite{Hager88,Dennis83}. For the equation~$z^3 - 1 =
0$, the Armijo rule modifies the Newton algorithm
(\ref{eq:Newton-mapping}) by introducing a fractional
damping coefficient~$1/2^j$ in front of the Newton step
as follows:
\begin{equation}
  \label{eq:Armijo} 
   z_{i+1} = z_i  -
     {1 \over 2^j} 
     \left( z_i^3 - 1 \over 3 z_i^2 \right) ,
\end{equation}
where~$j$ is a nonnegative integer.  Before the
approximate root~$z_i$ is corrected to give a new and
hopefully better estimate~$z_{i+1}$, the integer~$j$ is
increased from zero (i.e., the fraction of the step
added is decreased) until the following criterion is
just satisfied:
\begin{equation}
  \label{eq:Armijo-convergence} 
  \left| z_{i+1}^3 - 1 \right| 
  \leq
  \left( 1 - {1 \over 2^{j+1}} \right) |z_i^3 - 1 | .
\end{equation}
At this point, $z_{i+1}$ is computed from
\Eqref{eq:Armijo} with the given value of~$j$ and the
Newton method is iterated until convergence or some
maximum limit of iterations is attained. (If the
latter, then usually a new initial condition must be
tried.) Under certain assumptions of smoothness, one
can prove that the Armijo method is guaranteed to
converge to some root \cite{Dennis83}.
\begin{figure}[ht] 
  \centering
  \includegraphics[width=3.5in]{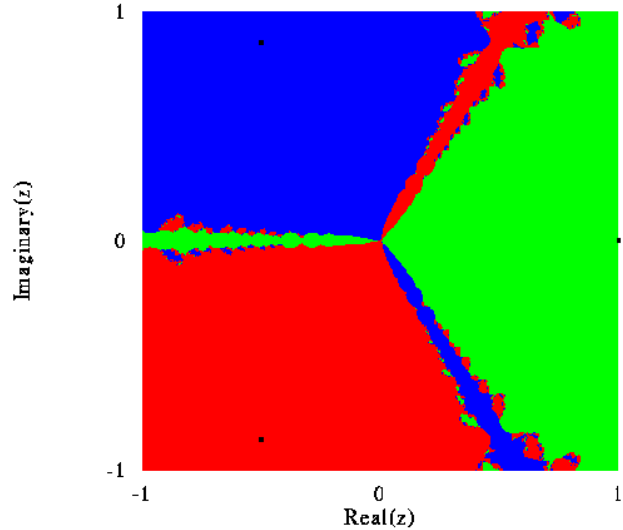}
\caption{Plot of the basins of attraction of the roots of 
  the equation~$z^3 - 1 = 0$ under Armijo's rule,
  Eqs.~(\ref{eq:Armijo})
  and~(\ref{eq:Armijo-convergence}), calculated in the
  same way as described in \Figref{fig:Newton}. The
  fractal structure is preserved and slightly modified
  by damping when compared with
  \Figref{fig:Newton}. The basin boundaries have a
  fractal dimension~$D \approx 1.29$ with a standard
  deviation of~$0.02$ as determined from a
  least-squares fit to a plot of~$\log(P_\varepsilon)$
  versus~$\log(1/\varepsilon)$ (not shown). }
\label{fig:Armijo} 
\end{figure}

A comparison of the basins of attraction of the
Newton-Armijo algorithm (\Figref{fig:Armijo}) with
those of the Newton algorithm (\Figref{fig:Newton})
yields several interesting insights.  First, the basin
boundaries are not smoothed out under damping but
retain a fractal structure.  Second, although the
basins themselves are substantially altered (there are
larger contiguous regions of a given color in
\Figref{fig:Armijo} compared to \Figref{fig:Newton}),
damping causes only a modest change in the basin
boundaries\footnote{A boundary was obtained by first
  separating the basin of attraction from the other
  basins, then by eliminating all the interior points,
  and finally, by computing the average pointwise
  dimension of the set of points that remained.} as
quantified by the fractal dimension~$D$.  The dimension
of the fractal boundaries in \Figref{fig:Armijo} is~$D
\approx 1.29$ whereas the fractal dimension in
\Figref{fig:Newton} has the somewhat larger value~1.42.
The approximately ten percent relative decrease in the
dimension seems consistent with the expected smoothing
effect arising from the Armijo factor $1 / 2^j$ at each
Newton-like step. Nevertheless, the dimension of the
basin boundary does not generally decrease under
damping as we point out in the next section.

\section{A Less Symmetric Example}

Our discussion for the polynomial~$z^3 - 1 = 0$ is
somewhat atypical in that there are only three roots
and the corresponding basins of attraction are related
by symmetry rotations around~$z=0$. To explore a more
representative example and to suggest the rich
geometric changes that damping can induce, we have also
studied numerically the approximate basins of
attraction of Newton and Newton-Armijo algorithms for
the following two-dimensional nonlinear system:
\begin{eqnarray} 
 { x - \sin(x) \cosh(y) } &=& {0 ,}\nonumber\\[-1.5ex]
 \label{eq:Complex} \\[-1.5ex]
 { y - \cos(x) \sinh(y) } &=& {0 ,}\nonumber
\end{eqnarray}
which the second author first came across in a
collection of numerical analysis problems~\cite[page
559, problem 16.86]{Scheid90}.
\begin{figure}[ht]  
  \centering
  \includegraphics[width=3.5in]{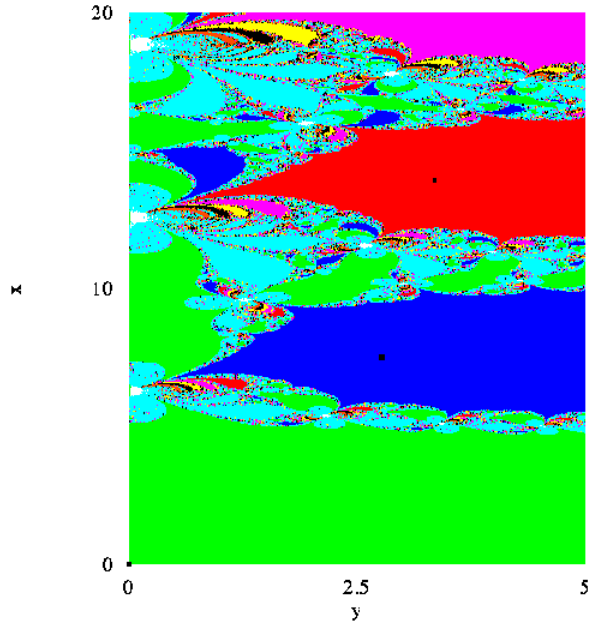}
\caption{Plot of the basins of attraction of the
  solutions of the system of equations
  (\ref{eq:Complex}) under Newton's method. Seven of
  the colors (green, purple, red, magenta, orange,
  yellow, and black) represent seven different basins of
  attraction (the most spatially extended regions). The blue
  regions represents all other basins while the white
  regions represent starting points for which the
  iterative process did not converge within
  1000~iterations. The figure was calculated using a
  mesh of $1500 \times 1500$ pixels spanning the
  spatial region $0 \le x \le 20$ and~$0 \le y \le 5$.
  The three solutions that lie inside this region are
  shown by the small black squares and have approximate
  coordinates $(x,y) = (0,0)$, $(7.5,2.8)$,
  and~$(13.9,3.4)$ respectively.  The fractal dimension
  of the boundary of the magenta-colored basin
  (belonging to the approximate solution~$(x,y)=(20.2,3.7)$)
  is~$D \approx 1.38 \pm 0.04$ as determined from a
  least-squares fit to a plot of~$\log(P_\varepsilon)$
  versus~$\log(1/\varepsilon)$.}
\label{fig:C_Newton} 
\end{figure}
\begin{figure}[ht]  
  \centering
  \includegraphics[width=3.5in]{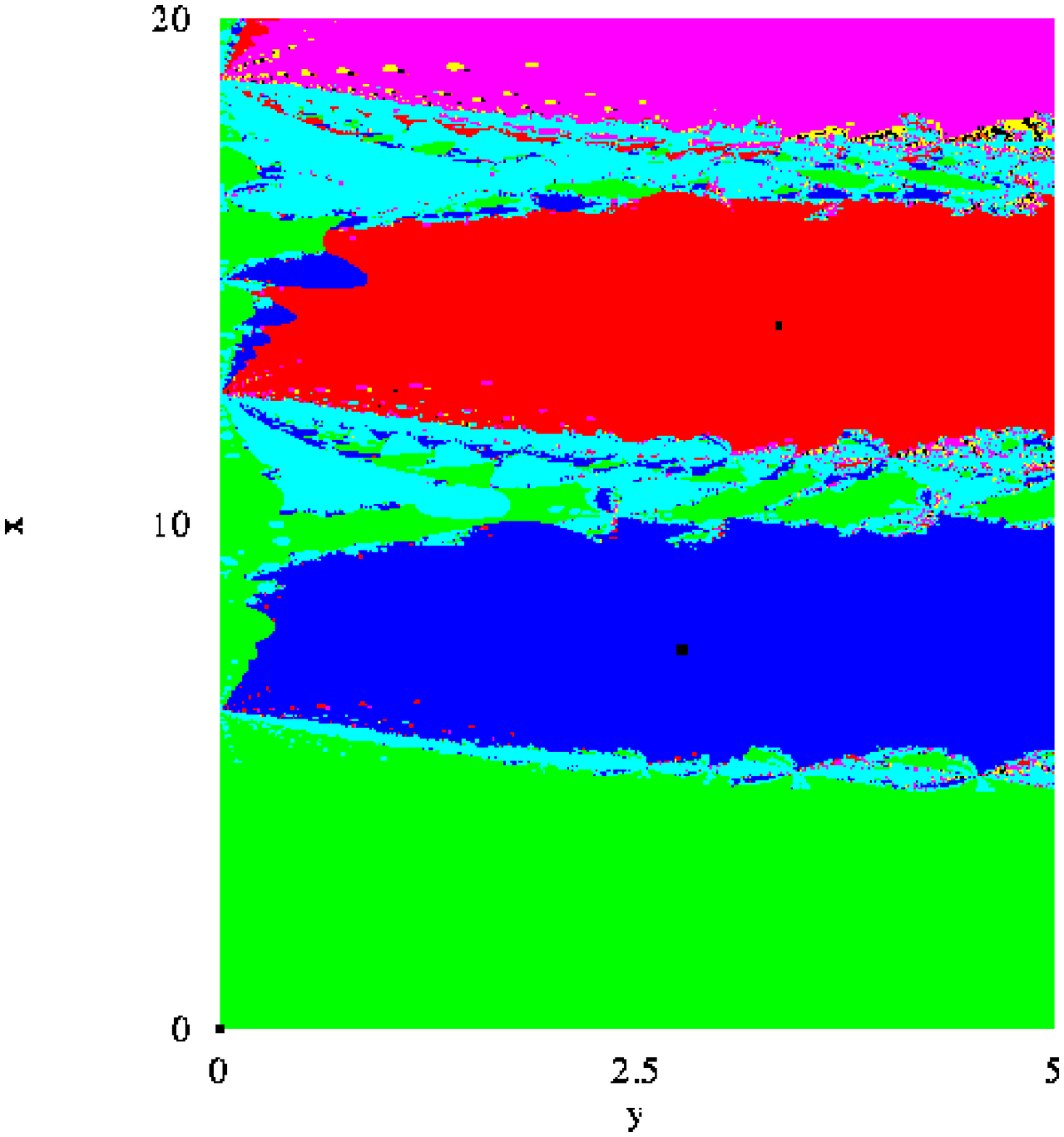}
\caption{The approximate basins of attraction of the
  solutions of \Eqref{eq:Complex} under Armijo's rule,
  using the same coloring scheme, resolution, and
  convergence criteria as in \Figref{fig:C_Newton}.
  The fractal dimension of the boundary of the
  magenta-colored basin colored is unchanged within
  numerical accuracy.}
\label{fig:C_Armijo} 
\end{figure}

Figures~\ref{fig:C_Newton} and~\ref{fig:C_Armijo} reveal a
dramatic change in the shape and number of basins of
attraction under damping although, within our numerical
accuracy, the fractal dimension of a particular basin
(the magenta basin belonging to the root~$(x,y)=(20.2,3.7)$)
is unchanged with a value~$D \approx 1.4$.

\begin{figure}[ht]   
  \centering
  \includegraphics[width=3.5in]{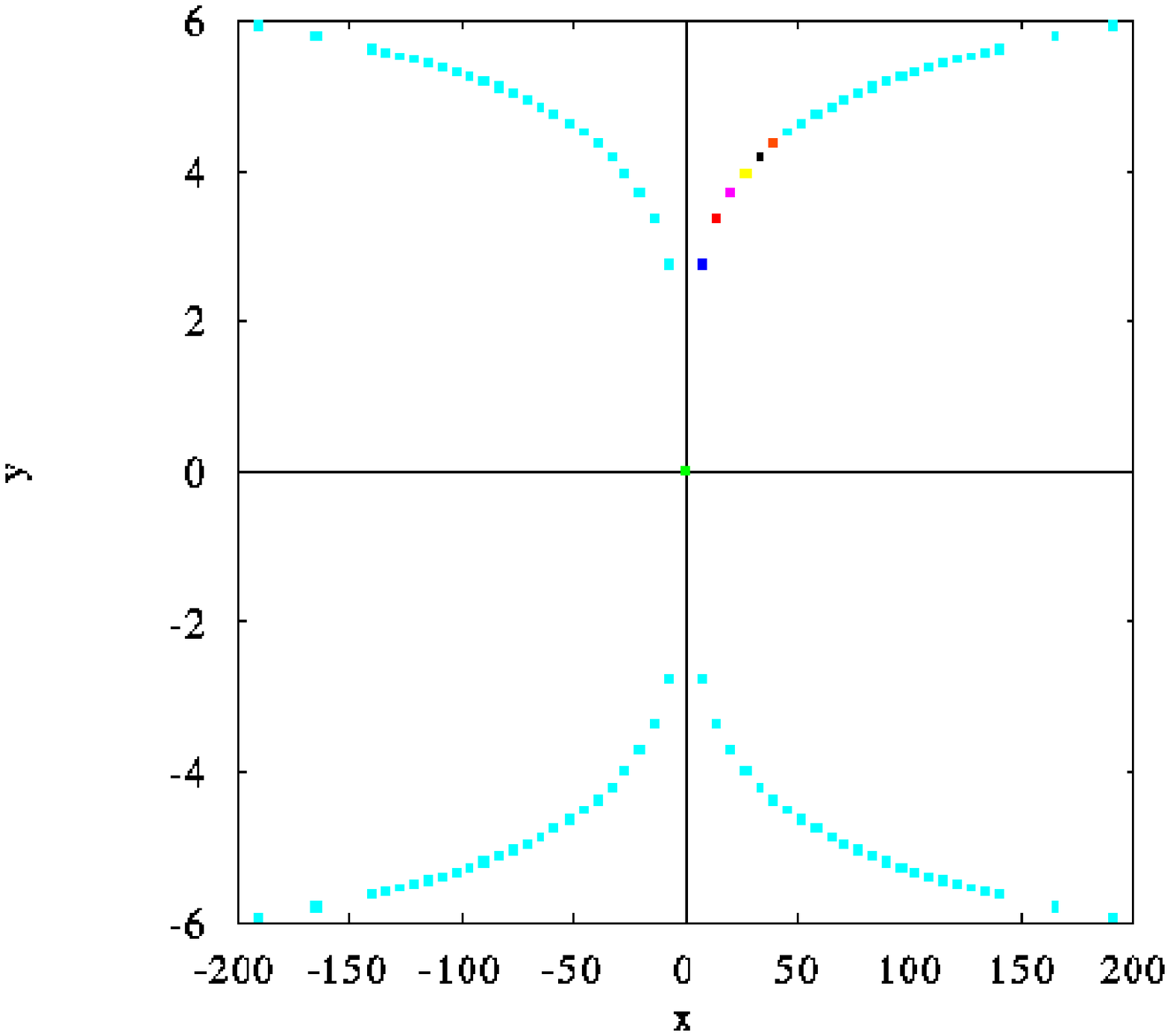}
\caption{Plot of some of the solutions of the system of 
  equations (\ref{eq:Complex}) under Newton's method, 
  using the same coloring scheme, resolution, and
  convergence criteria as in \Figref{fig:C_Newton}.
  All the solutions obtained under Armijo's rule were also
  obtained undere Newton's method; the basins of the
  solutions are nevertheless different. The color
  coding of the roots agrees with the color of the
  basins in \Figref{fig:C_Newton} and \Figref{fig:C_Armijo}.}
\label{fig:C_Roots} 
\end{figure}

Using starting points centered on $1500 \times 1500$
pixels spanning the region $0 \le x \le 20$ and~$0 \le
y \le 5$, we found that Newton's method
converged\footnote{The criterion for convergence was
  that the magnitude of the residual was smaller
  than~$10^{-12}$ and the magnitude of the correction
  term was smaller than~$10^{-4}$.} towards~$120$
different solutions\footnote{Two solutions were
  considered different if the Euler distance between
  them in the $x$-$y$ plane was greater than
  $10^{-3}$.} Some of these solutions are shown
in~\Figref{fig:C_Roots} and are color coded to indicate
the associated basin of attraction. In contrast, the
same initial conditions with Armijo's rule yielded
only~$53$ different solutions, a substantially smaller
number. These solutions were a subset of the 120 solutions 
obtained using Newton's method.

Figures~\ref{fig:C_Newton}-\ref{fig:C_Roots} raise
fascinating open questions in addition to those already
discussed above in the context of \Figref{fig:Newton}.
As an example, what has happened to the many basins of
attraction that are now missing in the damped algorithm
(53 solutions versus 120 for initial conditions in a
certain given region)?  Does the Wada property still
apply to the various basins and, if so, which basins
are linked by a given boundary? If not all basins are
linked by the Wada property, do the different basin
boundaries have different fractal dimensions and, if
so, what determines these different dimensions?

Finally, one can again ask how the geometry in
\Figref{fig:C_Armijo} might depend on the damping
algorithm. A canonical and interesting choice to study
would be {\em continuous} damping in the sense of using
an infinitesimal damping factor~$ds$ along a Newton
direction.  (We are indebted to our colleague, Dr.\ 
Donald Rose, for this suggestion.) Instead of iterating
an Armijo-Newton map similar to Eqs.~(\ref{eq:Armijo})
and~(\ref{eq:Armijo-convergence}), one would then
integrate a system of~$N$ ordinary differential
equations given by:
\begin{equation}
  { d{\bf X} \over ds}  =
    - {\bf J}^{-1} \cdot {\bf F}({\bf X}) ,
\label{eq:Continuous}
\end{equation}
where the~$N$-dimensional system of equations is given
by ${\bf F}({\bf X}) = 0$, where~${\bf J}({\bf X})$ is
the~$N \times N$ Jacobian matrix~${\bf J}({\bf X}) =
\partial{\bf F}/\partial{\bf X}$, and where~$s$ is some
arbitrary continuous parameter. Starting from some
initial guess~${\bf X}_0$, numerical integration of
\Eqref{eq:Continuous} will generate a path ${\bf X}(s)$
that must end up at a solution provided the Jacobian
matrix remains nonsingular along this path.
Challenging questions are raised in this continuous
case: how long one must integrate to get close to a
root?  What integration accuracy is needed? What
numerical algorithms should be used? What are the
basins of attraction and are they fractal?

\section{Conclusions}

Numerical analysis is an exciting and attractive
subject specifically because it is fairly easy to to
investigate novel problems even at the level of an
introductory course. As an illustration of this, we
have investigated numerically and graphically the
problem of how a simple damping algorithm for Newton's
method, Armijo's rule, modifies the fractal basins of
attraction found for the complex polynomial~$z^3 - 1 =
0$ . We have also investigated a less symmetric and
perhaps more general case, \Eqref{eq:Complex}, in which
the basins of attraction are fundamentally changed upon
damping. Our goal here was not to present a detailed or
complete analysis but instead to suggest how students
learning numerical analysis can easily find and explore
new problems. The field of nonlinear dynamics
\cite{Strogatz94} raises many related questions of
geometric structure and their relation to numerical
algorithms and is a valuable source of other ideas of
pedagogical value for courses in numerical analysis.


\end{document}